\documentclass[amsmath,amssymb,showpacs,draft,preprint,floatfix]{revtex4}
\usepackage{graphicx}
\usepackage{latexsym}
\usepackage{amsmath}
\usepackage{amsfonts}
\usepackage{amssymb}
\begin{document}

\title{Squeeze operators in classical scenarios}
\author{Jorge A. Anaya-Contreras{$^1$}, Arturo Z\'u\~niga-Segundo{$^1$}, Francisco Soto-Eguibar{$^2$}, V\'ictor Arriz\'on{$^2$}, H\'ector M. Moya-Cessa{$^2$}}

\address{$^1$Departamento de F\'isica, Escuela Superior de F\'isica y Matem\'aticas, IPN\\
Edificio 9 Unidad Profesional `Adolfo L\'opez Mateos', 07738 M\'exico D.F., Mexico\\
$^2$Instituto Nacional de Astrof\'isica \'Optica y Electr\'onica\\
Calle Luis Enrique Erro No. 1, Sta. Ma. Tonantzintla, Pue. CP 72840, Mexico}
\begin{abstract}
 We analyse the paraxial field propagation in the realm of classical optics, showing that it can be written as the action of the fractional Fourier transform, followed by the squeeze operator applied to the initial field. Secondly, we show that a wavelet transform may be viewed as the application of a displacement and squeeze operator onto the mother wavelet function.
\end{abstract}
\maketitle

\section{Introduction} \label{sec:intro}
In the late seventies, squeezed states were introduced \cite{Loudon,Barnett}. On the one hand, Yuen \cite{Yuen} defined them squeezing the vacuum and then displacing the resulting state. On the other hand, Caves \cite{Caves} defined them by displacing the vacuum and then squeezing the produced coherent state. Squeezed states have been shown to produce ringing revivals (a fingerprint that a squeezed state is used) in the interaction between light and matter \cite{Squeezed}. Applications of quantum techniques in classical optics have been the subject of many studies during the last years \cite{Tailoring,W-state}.  Along the same line, one of the goals of this article is to show, that in a mathematical sense, the squeeze operator could have been introduced in the description of free light  propagation, i.e. in the domain of classical optics, at least a hundred years earlier. We also show that we can use such squeeze operators to write the continuous wavelet transform as its average with the mother wavelet function, a displacement operator and the function to be transformed.

\section{Squeezed states}
As we already explained in the introduction, there are two equivalent forms to define the squeezed states. In the first one, introduced by Yuen \cite{Yuen}, squeezed states are obtained from the vacuum as
\begin{equation}
|\alpha ;r\rangle =\hat{S}(r)\hat{D}(\alpha )|0\rangle =\hat{S}(r)|\alpha \rangle ,
\end{equation}
where
\begin{equation}\label{sqop}
\hat{S}(r)=\exp \left[\frac{r}{2}\left(\hat{a}^2-\hat{a}^{\dagger^{2}}\right)\right]
\end{equation}
is the queeze operator and $\hat{D}(\alpha )$ is the Glauber displacement operator \cite{Glauber}. Here $\hat{a}= \frac{1}{\sqrt{2}}(x+d/dx)$ and $\hat{a}^{\dagger}= \frac{1}{\sqrt{2}}(x-d/dx)$ are the ladder operators \cite{Arfken}. In this view, squeezed states are created displacing the vacuum, and after, squeezing it. Note that when the squeeze parameter $r$ is set to zero, the squeezed states reduce to the coherent states. In this work, we will consider only real squeeze parameters, as that is enough for our intentions.\\
In the definition of the squeezed states followed by Caves \cite{Caves}, the vacuum is squeezed and the resulting state is then displaced; which means  that in this approach, they are given by the expression
\begin{equation} \label{SS}
|\alpha ';r'\rangle =\hat{D}\left(\alpha '\right)\hat{S}\left(r'\right)|0\rangle .
\end{equation}
Both definitions of the squeezed states agree when the squeeze factor is the same, \(r'=r\), and when the modified amplitude \(\alpha '\) of the Caves approach is given by
\begin{equation}
\alpha '=\mu \alpha -\nu \alpha^*,
\end{equation}
being
\begin{equation}
\mu =\cosh \, r
\end{equation}
and
\begin{equation}
\nu =\sinh \, r.
\end{equation}
To analyse the uncertainties in the position and in the momentum of the squeezed states, we introduce, following Loudon and Knight \cite{Loudon}, the quadrature
operators
\begin{equation}
\hat{X}=\frac{\hat{a}+\hat{a}^{\dagger }}{2}=\frac{\hat{x}}{\sqrt{2}}
\end{equation}
and
\begin{equation}
\hat{Y}=\frac{\hat{a}-\hat{a}^{\dagger }}{2i}=\frac{\hat{p}}{\sqrt{2}},
\end{equation}
where \(\hat{x}\) is the position operator, \(\hat{p}\) is the momentum operator and the  operators $\hat{a}$ and $\hat{a}^{\dagger}$ are the annihilation and creation operators of the harmonic oscillator, respectively.  Note that the quadrature operators are essentially the position and momentum operators; this definition just provides us with two operators that have the same dimensions.\\
In order to show that really the squeezed states are minimum uncertainty states, we need to calculate the expected values in the squeezed state (1) of the quadrature operators (7) and (8), and its squares. Using (7) and (1), we obtain
\begin{equation}
\langle \alpha ;r|\hat{X}|\alpha ;r\rangle =\langle \alpha |\hat{S}^{\dagger }(r)\frac{\hat{a}+\hat{a}^{\dagger }}{2}\hat{S}(r)|\alpha\rangle .
\end{equation}
The action of the squeeze operator on the creation and annihilation operators is obtained using the Hadamard's lemma \cite{nos2,miller,hall},
\begin{equation}
\hat{S}^{\dagger }(r)\hat{a}\hat{S}(r)=\mu \hat{a}-\nu \hat{a}^{\dagger },\text{          }\hat{S}^{\dagger }(r)\hat{a}^{\dagger }\hat{S}(r)=\mu \hat{a}^{\dagger }-\nu \hat{a},
\end{equation}
such that
\begin{equation} \label{action}
\hat{S}^{\dagger }(r)\frac{\hat{a}+\hat{a}^{\dagger }}{2}\hat{S}(r)=e^{-r}\hat{X}.
\end{equation}
Therefore, as $\hat{a}|\alpha \rangle =\alpha |\alpha \rangle$ and \(\langle \alpha |\hat{a}^{\dagger }=\langle \alpha |\alpha ^*\), it is easy to see that
\begin{equation}
\langle \alpha ;r|\hat{X}|\alpha ;r\rangle =e^{-r}\frac{\alpha +\alpha ^*}{2},
\end{equation}
and that
\begin{equation}
\langle \alpha ;r|\hat{X}^2|\alpha ;r\rangle =e^{-2r}\frac{1+2\left| \alpha \right| ^2+\alpha ^2+\alpha ^{*^{2}}}{4}.
\end{equation}
So, we obtain for the uncertainty in the quadrature operator \(\hat{X}\),
\begin{equation}
\Delta \, X\equiv \sqrt{\langle \alpha ;r|\hat{X}^2|\alpha ;r\rangle -\langle \alpha ;r|\hat{X}|\alpha ;r\rangle ^2}=\frac{e^{-r}}{2}.
\end{equation}
Proceeding in exactly the same way for the quadrature operator \(\hat{Y}\), we obtain
\begin{equation}
\Delta \, Y\equiv \sqrt{\langle \alpha ;r|\hat{Y}^2|\alpha ;r\rangle -\langle \alpha ;r|\hat{Y}|\alpha ;r\rangle ^2}=\frac{e^r}{2}.
\end{equation}
As we already said, we can then think in the position eigenstates and in the momentum eigenstates as limiting cases of squeezed states. Indeed, when the squeeze parameter \(r\) goes to infinity, the uncertainty in the position goes to zero, and the momentum is completely undetermined. Of course, when the squeeze parameter goes to minus infinity, we have the inverse situation, and we can define in that way the momentum eigenstates. In the two following sections, we use the Yuen and the Caves definitions of the squeezed states to test this hypothesis.

\section{Squeeze and fractional Fourier operators in paraxial optics}
The propagation of light in free space can be described by the paraxial equation
\begin{equation}\label{Paraxial}
i\frac{\partial E(x,y,z)}{\partial z}= -\frac{1}{2}\frac{\partial^2 E(x,y,z)}{\partial x^2}-\frac{1}{2}\frac{\partial^2 E(x,y,z)}{\partial y^2},
\end{equation}
where we have set the wavevector $k$ equal to one. We define $\hat{p}_{\alpha}=-i\partial/ \partial \alpha$, with $\alpha=x,y$ such that we rewrite the above equation as (we obviate the variables $x$ and $y$)
\begin{equation}\label{Paraxial2}
i\frac{\partial E(z)}{\partial z}= \frac{\hat{p_x}^2+\hat{p_y}^2}{2} E(z),
\end{equation}
that allows to give the simple formal solution
\begin{equation}\label{Sol}
E(z)=\exp\left[ -i\frac{z}{2}(\hat{p_x}^{2}+\hat{p_y}^{2})\right] E(0).
\end{equation}
We use the annihilation and creation operators for the harmonic oscillator,
\begin{equation}\label{PCC2}
\hat{a}_{\alpha}=\frac{\hat{\alpha}+i\hat{p}_{\alpha}}{\sqrt{2}}, \qquad \hat{a}_{\alpha}^{\dagger}=\frac{\hat{{\alpha}}-i\hat{p}_{\alpha}}{\sqrt{2}}, \qquad {\alpha}=x,y
\end{equation}
to cast  Eq. \eqref{Sol} into
\begin{eqnarray}\label{Sol2}
E(z)&=&\exp\left[ -i\frac{z}{2}\left( \hat{n}_x+\frac{1}{2}-\frac{\hat{a}_x^2}{2}-\frac{\hat{a}_x^{\dagger 2}}{2}\right) \right]\\ \nonumber &\times& \exp\left[ -i\frac{z}{2}\left( \hat{n}_y+\frac{1}{2}-\frac{\hat{a}_y^2}{2}-\frac{\hat{a}_y^{\dagger 2}}{2}\right) \right]
E(0),
\end{eqnarray}
being $\hat{n}_{\alpha}=\hat{a}_{\alpha}^{\dagger }\hat{a}_{\alpha}$ the so-called number operator in quantum optics. In the following, we show how to factorize this exponential as the product of a squeeze and a fractional Fourier transform operators\cite{Namias}.

\subsection{Evolution operator factorization}
Each exponential in \eqref{Sol2} may be written as
\begin{equation}\label{PCC3}
\exp\left[ -i\frac{z}{2}\left(2\hat{K}_{0}-\hat{K}_{+}-\hat{K}_{-}\right)\right],
\end{equation}
with (for simplicity, we drop the $\alpha$ subindexes of the annihilation and creation operators)
\begin{equation}\label{PCC4}
\hat{K}_{0}=\frac{1}{2}\left(\hat{a}^{\dagger}\hat{a}+\frac{1}{2}\right), \qquad \hat{K}_{+}=\frac{\hat{a}^{\dagger \, 2}}{2}, \qquad \hat{K}_{-}=\frac{\hat{a}^{2}}{2},
\end{equation}
which are the elements of a  Lie Algebra $su(1,1)$ \cite{hall} and satisfy the following commutation relations
\begin{equation}\label{PCC5}
\left[\hat{K}_{-},\hat{K}_{+}\right]=2\hat{K}_{0}, \qquad \left[\hat{K}_{0}, \hat{K}_{+}\right] = \hat{K}_{+}, \qquad
\left[\hat{K}_{0},\hat{K}_{-}\right] =-\hat{K}_{-}.
\end{equation}
According to Fisher {\it et al.} \cite{Nieto} the generators of the $su(1,1)$ algebra admit the matrix representation
\begin{equation}\label{PCC6}
\hat{K}_{0}=
\left(\begin{array}{ll}1/2 & 0\\
0 & -1/2
\end{array}\right)
\end{equation}
and
\begin{equation}\label{PCC6.1}
\hat{K}_{+}=
\left(\begin{array}{ll}
0 & 1\\
0 & 0
\end{array}\right), \qquad
\hat{K}_{-}=
\left(\begin{array}{ll}
0 & 0\\
-1 & 0
\end{array}\right),
\end{equation}
such that Eq. \eqref{PCC3} may be rewritten as
\begin{equation}\label{PCC7}
\exp\left[ {-i\frac{z}{2}\left(2\hat{K}_{0}-\hat{K}_{+}-\hat{K}_{-}\right)}\right]  =\left(\begin{array}{ll} 1-i\frac{z}{2}& i\frac{z}{2} \\ -i\frac{z}{2} & 1+i\frac{z}{2}\end{array}\right),
\end{equation}
because
\begin{equation}\label{PCC8}
\left(\begin{array}{ll} 1&-1\\ 1&-1 \end{array}\right)^{n}=0  \qquad \mathrm{for \; all}   \qquad n = 2,3, \dots .
\end{equation}
We now assume that exist two numbers $\xi = r e^{i\theta}$ and $ \omega $ such that
\begin{align}\label{PCC9}
\left(\begin{array}{ll} 1-i\frac{z}{2}& i\frac{z}{2}
\nonumber \\
-i\frac{z}{2} & 1+i\frac{z}{2}\end{array}\right) &= \exp\left[ -i\frac{z}{2}\left(2\hat{K}_{0}-\hat{K}_{+}-\hat{K}_{-}\right)\right]
\nonumber \\
&= \exp(-i2\omega \hat{K}_{0})\exp\left(\xi \hat{K}_{+}-\xi^{*}\hat{K}_{-}\right)
\nonumber \\
&= \left(\begin{array}{ll}e^{-i\omega}\cosh r & e^{i(\theta-\omega)}\sinh r\\ e^{i(\omega-\theta)}\sinh r & e^{i\omega} \cosh r\end{array}\right),
\end{align}
with
\begin{equation}\label{PCC10}
1+i\frac{z}{2}= e^{i\omega} \cosh r,\qquad -i\frac{z}{2}=e^{-i\theta}e^{i\omega}\sinh r,
\end{equation}
or
\begin{equation}\label{PCC11}
e^{i\omega}=\frac{1+i\frac{z}{2}}{\sqrt{1+\left(\frac{z}{2}\right)^{2}}}, \qquad e^{i\theta}=ie^{i\omega},
\end{equation}
and
\begin{equation}
r=\ln \left({\sqrt{1+\left(\frac{z}{2}\right)^{2}}-\frac{z}{2}}\right).
\end{equation}
Therefore, we may write
\begin{eqnarray}\label{PCC12}
\exp\left(-i\frac{z}{2} \hat{p}^{2}\right)&=&\hat{S}\left(i r e^{-i\omega}\right)\exp \left[ -i\omega \left(\hat{a}^{\dagger}\hat{a}+\frac{1}{2}\right)\right] \\ \nonumber &=&\hat{S}\left(\xi \right)\hat{F}(\omega),
\end{eqnarray}
where $\hat{S}(\xi)$ is the squeeze operator \cite{Caves,Yuen}, Eq. \eqref{sqop}, and  $\hat{F}(\omega)$ is the fractional  Fourier transform (see for instance Namias \cite{Namias}), with
\begin{equation}\label{PCC14}
\xi= i r e^{- i\omega}.
\end{equation}
Then the solution to the paraxial wave equation reads
\begin{equation}\label{Solfin}
E(z)=\hat{S}_x\left(\xi \right)\hat{S}_y\left(\xi \right)\hat{F}_x(\omega)\hat{F}_y(\omega)
E(0),
\end{equation}
that is nothing but the application of squeeze operators applied to the two-dimensional fractional Fourier transform of the field at $z=0$. It is not difficult to show that for large $z$, $\omega \rightarrow \pi/2$ such that $\xi \rightarrow r$ and the fractional Fourier transform  becomes the (complete) Fourier transform. The solution to the paraxial equation for $z$ large therefore reads
\begin{equation}\label{SolFin}
E(x,y,z)=\hat{S}_x(r)\hat{S}_y(r)\tilde{E}(x,y,0),
\end{equation}
with $\tilde{E}(x,y,0)$ the two-dimensional Fourier transform of $E(x,y,0)$. Further application of the squeeze operator yields
\begin{equation}\label{SolFinal}
E(x,y,z)=e^{-r}\tilde{E}(xe^{-r},ye^{-r},0).
\end{equation}
As can also be shown from Eqs.\eqref{PCC11}, when $z$ is very large $r \rightarrow \ln(z)$, thus
\begin{equation}\label{SolFinal2}
E(x,y,z)=\frac{1}{{z}}\tilde{E}\left(\frac{x}{z},\frac{y}{z},0\right),
\end{equation}
which, up to a phase, is the expected expression \cite{pellat,akhmanov,vic1}.

\section{Wavelet transforms}
The integral (continuous) wavelet transform of a function $f(x)$ is given by \cite{meyer,charles,Wavelets}
\begin{equation}\label{wavel0}
{\cal F}(a,b)=\frac{1}{\sqrt{|a|}}\int_{-\infty}^{\infty} \psi^*\left(\frac{x-b}{a}\right)f(x)dx,
\end{equation}
where $\psi(x)$ is the so called mother wavelet function. Because $\exp(ib\hat{p})g(x)=g(x+b)$ the above equation may be written in the form
\begin{equation}\label{wavel1}
{\cal F}(a,b)=\frac{1}{\sqrt{|a|}}\int_{-\infty}^{\infty} \psi^*\left(\frac{x}{a}\right) \exp(ib\hat{p}) f(x)dx,
\end{equation}
and using the squeeze operator introduced above and the equations presented in the Appendix, we may write in Dirac notation the simple form
\begin{equation}\label{wavel2}
{\cal F}(a,b)=\langle\psi|\hat{S}^{\dagger}(r) \exp(ib\hat{p})|f\rangle .
\end{equation}
If we choose the very simple mother wavelet function, namely the state $|0\rangle$, i.e., the Hermite-Gaussian $\psi_0(x)$, the wavelet integral transform reduces to \cite{To-Ve,chinos}
\begin{equation}\label{wavel4}
{\cal F}(a,b)=\langle b,r|f\rangle .
\end{equation}
where $|b,r\rangle$ has the form of a squeezed state, equation (\ref{SS}).

\section{Conclusions}
We have shown that some techniques that are common in quantum mechanics may be applied in classical scenarios used in optics. In particular, we have written the  free propagation of a field as the application of the product of squeeze operators corresponding to the variables $x$ and $y$ and the two-dimensional Fractional Fourier operator to the field at $z=0$. Finally we showed that it is possible to write the continuous wavelet transform as the application of a "bra" mother wavelet to a "ket" that corresponds to the function to be transformed.

\section*{Appendix}
A function $\psi(x)$ may be expanded in Hermite-Gaussian functions as
\begin{equation}
\psi(x)=\sum_{n=0}^{\infty}c_n\psi_n(x),
\end{equation}
with
\begin{equation}
\psi_n(x)=\frac{1}{\sqrt{2^n n! \sqrt{\pi}}} \exp(-\frac{x^2}{2}) H_n(x),
\end{equation}
where $H_n(x)$ are the Hermite polynomials of order $n$. The coefficients $c_n$ may be calculated from the integral $c_n=\int_{-\infty}^{\infty}\psi(x)\psi_n(x) dx$.
In Dirac notation the above may be casted as
\begin{equation}|\psi\rangle=\sum_{n=0}^{\infty}c_n|n\rangle,
\end{equation}
where the states $|n\rangle$ are the number or Fock states. The coefficients $c_n$ are calculated by the quantity $c_n=\langle n|\psi\rangle$.
If we apply a squeeze operator to the function $\psi(x)$ we obtain
\begin{equation}
S(r)\psi(x)1=S(r)\psi(x)S^{\dagger}(r)S(r)1,
\end{equation}
where we have multiplied by $1$ and we have introduced an extra $1$, namely $S^{\dagger}(r)S(r)$, in the right hand side of the equation above. From \eqref{action}, we can see that $S(r)\psi(x)S^{\dagger}(r)=\psi(xe^{-r})$ and the action of $S(r)$ on $1$ is
\begin{align}
S(r)1&=\exp\left[ -i\frac{r}{2}(x\hat{p}+\hat{p}x)\right] 1
\\ \nonumber &
=\exp\left[ i\frac{r}{2}(2x\hat{p}-i)\right] 1=\exp(-r/2).
\end{align}
Now, the integral of two functions is given by
\begin{equation}\label{integra}
\int_{-\infty}^{\infty} \psi^*(x)f(x)dx=\sum_{n=0}^{\infty}\sum_{m=0}^{\infty}\int_{-\infty}^{\infty} c^{(\psi) *}_nc^{(f)}_m \psi_n(x)\psi_m(x),
\end{equation}
where we have made explicit that the coefficients are related to specific functions. Because of orthogonality it reduces to
\begin{equation}\label{integral}
\int_{-\infty}^{\infty} \psi^*(x)f(x)dx=\sum_{n=0}^{\infty} c^{(\psi) *}_nc^{(f)}_n=\langle \psi|f\rangle.
\end{equation}


\begin{thebibliography}{99}
\bibitem{Loudon} R.Loudon and P.L. Knight, Squeezed light. J. Mod. Opt. {\bf 34}, 709-759 (1987).
\bibitem{Barnett} S.M. Barnett, A. Beige, A. Ekert, B.M. Garraway, {\it et al.}, Journeys from quantum optics to quantum technology. {Progress in Quantum Electronics} {\bf 54}, 19-45 (2017).
\bibitem{Yuen} H.P. Yuen, Two-photon coherent states of the radiation field. Phys. Rev. A {\bf 13}, 2226-2243 (1976).
\bibitem{Caves} C.M. Caves, Quantum-mechanical noise in an interferometer. Phys. Rev. D {\bf 23},  1693-1708 (1981).
\bibitem{Squeezed} H. Moya-Cessa and A. Vidiella-Barranco, Interaction of squeezed light with two-level atoms. J. Mod. Optics {\bf 9},  2481-2499 (1995).
\bibitem{Tailoring} A. Perez-Leija, R. Keil, A. Szameit, A.F. Abouraddy, H. Moya-Cessa, D.N. Christodoulides, Tailoring the correlation and anticorrelation behavior of path-entangled photons in Glauber-Fock oscillator lattices, Phys. Rev. A {\bf 85}, 013848 (2012).
\bibitem{W-state} A. Perez-Leija, J.C. Hernandez-Herrejon, H. Moya-Cessa, A. Szameit and  D.N. Christodoulides, Generating photon-encoded W states in multiport waveguide-array systems. Phys. Rev. A {\bf 87}, 013842 (2013).
\bibitem{Glauber}  R. J. Glauber, Coherent and
Incoherent States of the Radiation Field. Phys.\ Rev.\ Lett.\/ {\bf 10}, 84 (1963).
\bibitem{Arfken} G. Arfken, {\it Mathematical methods for physicists,}
(Academic Press, Inc., 3rd Edition, 1985).
\bibitem{nos2} H.M. Moya-Cessa and F. Soto-Eguibar, {Introduction to Quantum Optics}, Rinton Press Inc., 2011.
\bibitem{miller} W. Miller, {Symmetry Groups and their Applications}, Academic Press, New York, 1972.
\bibitem{hall} B. C. Hall, {Lie Groups, Lie Algebras, and Representations. An Elementary Introduction}, Graduate Texts in Mathematics, 222 (2nd ed.), Springer, 2015.
\bibitem{Namias} V. Namias, The Fractional order Fourier transform and its application to quantum mechanics. J. Inst. Maths. Applics. {\bf 25}, 241-265 (1980).
\bibitem{Nieto} R.A. Fisher, M.M.  Nieto, and  V.D. Snadberg, Impossibility of naively generalizing squeezed coherent states. Phys. Rev. D  {\bf 29}, 1107-1110 (1984).

\bibitem{pellat} P. Pellat-Finet, Optique de Fourier. Th\'eorie m\'etaxiale et fractionaire, Springer, 2009
\bibitem{akhmanov} S. A. Akhmanov and S. Y. Nikitin, Physical Optics, Clarendon Press, Oxford 1997.
\bibitem{vic1} V. Arriz\'on, F. Soto-Eguibar, D. S\'anchez-De-La-Llave, and H. M. Moya-Cessa,  Conversion of any finite bandwidth optical field into a shape invariant beam. OSA Continuum {\bf 1},  604-612 (2018).
\bibitem{meyer} Y. Meyer, {Wavelets and Operators}. Cambridge University Press 1992. 

\bibitem{charles} Charles K., {An Introduction to Wavelets}. Academic Press 1992.
\bibitem{Wavelets} P.S. Addison, The illustrated wavelet transform handbook: introductory theory and applications in science, engineering, medicine and finance, 2nd edition, CRC Press 2016.

\bibitem{To-Ve} Go. Torres-Vega and J.H. Frederick, A quantum mechanical representation in phase space. J.\ Chem.\ Phys.\/ {\bf 98}, 3103 (1993).

\bibitem{chinos} L.-y. Hu, H.-y. Fan, and H.-l. Lu, Explicit state vector for Torres-Vega-Frederick phase space representation and its statistical behavior. J.\ Chem.\ Phys.\/ {\bf 128}, 054101 (2008).


\end{thebibliography}
\end{document}